\documentclass[aps,preprint,tighten,floats,epsf,rotate]{revtex4}
\usepackage{graphicx}

\begin{document}
\title{Neutrino inflation of baryon inhomogeneities in strong magnetic fields.}
\author{Soma Sanyal \footnote{e-mail: soma@iucf.indiana.edu}}
\affiliation{Dept. of Physics, Indiana University, Bloomington, IN 47405}

\begin{abstract}
Baryon inhomogeneitites formed in the early universe are important as 
they affect the nucleosynthesis calculations. Since they are formed much 
before the nucleosynthesis epoch, neutrino inflation plays a crucial role 
in damping out these fluctuations. Now neutrinos, in turn, are affected by magnetic 
fields which may be present in the early universe. 
In this work we study the evolution of baryonic inhomogeneities due to 
neutrino induced dissipative processes in the presence of a background 
magnetic field. We find that at higher temperatures the dissipation of the 
inhomogeneities are enhanced as the magnetic field increases. Our study also
shows that at lower temperatures the same magnetic field may produce 
less dissipation. Though we limit our study to temperatures below the 
quark-hadron transition we do establish that magnetic fields present in the
early universe  affect the dissipation of baryonic inhomogeneities.     
\end{abstract}

\maketitle

\section{Introduction}

Baryon inhomogeneities are formed during various epochs in the early universe
\cite{inhomo}.
These inhomogeneities are important as their presence during the 
nucleosynthesis epoch affects the neutron to proton ratio and 
thus the calculated abundances of the light elements. These baryon 
inhomogeneities get dissipated as the universe gets cooler by the 
combined effects of neutrino inflation (before 1 MeV) and baryon diffusion (
after neutrino decoupling). The baryon inhomogeneities have a higher baryon 
number density than the surrounding plasma, so to maintain pressure equilibrium 
with the plasma they have to have a lower temperature than 
the surrounding plasma.  As long as neutrinos are free streaming 
(over the lengthscale of the inhomogeneties) they pass from the high temperature 
plasma to the low 
temperature inhomogeneities and deposit heat in them. The deposition of heat 
disturbs the pressure balance and the inhomogeneity increases in volume
to achieve pressure equilibrium again. Thus the volume of the inhomogeneity 
increases while its amplitude goes down due to transfer of heat 
by the neutrinos. Detailed study of the 
dissipation  of these baryon fluctuations have been carried out 
\cite{jedamzik,heckler} previously. As seen in these studies neutrino inflation 
reduces the inhomogeneities but in most cases is unable 
to wipe them out completely. This is especially true if the inhomogeneity is
large. Smaller inhomogeneities are however wiped out by neutrino inflation.  

The studies done previously have been done without the presence of a 
background magnetic fields. There are ample evidences to show that the early 
universe had large magnetic fields \cite{kronberg}. Recent studies have also 
shown that neutrinos though uncharged are affected 
by magnetic fields especially in the early universe plasma \cite{kaushik}. 
Magnetic fields increase heat loss from the 
background plasma by the enhanced creation of neutrino pairs by e+e- 
annihilation, neutrino synchrotron radiation and other processes.
These neutrinos pass through the 
baryon inhomogeneties and increase the total heat deposited in them.
In this work, we study the effect of this on the baryon inhomogeneities. 

Heat loss from the background plasma by neutrinos in 
presence of magnetic field has been studied with respect to  neutron stars 
and supernovae explosions. These studies are thus for much lower temperatures. 
($T= 10^9 K$). Since neutrino inflation of baryon inhomogeneities occur at 
much higher temperatures($T>>10^9 K$), and  very few studies have been 
done for such temperatures; we believe that our work will also provide a 
motivation to carry out further studies of neutrino effects in higher 
temperature plasmas in the presence of magnetic fields. 

In section II we first review heat deposition in baryon inhomogeneities,
then in section III, we review neutrinos in high temperature
magnetic plasmas. In section IV, we present our calculations for the dissipation
of baryonic inhomogeneities, due to neutrinos in the presence of a magnetic 
field and give our results. In section V, we briefly discuss how we can 
get such strong magnetic fields before the nucleosynthesis epoch.
Section VI presents our conclusions and some discussions regarding the 
approximations considered in this work.

\section{Heat deposition in baryon inhomogeneities.}
Baryon inhomogeneities have a lower temperature than the surrounding plasma.
Free streaming particles moving from the plasma into these inhomogemeities 
tend to deposit heat in them. The plasma loses heat to these 
inhomogeneities continuously until they are completely wiped out.  
Since plasma heat loss mechanism is enhanced by magnetic field, the  
heat deposition in baryon inhomogeneities may get enhanced. This will cause 
the inhomogeneity to dissipate faster. We now review heat deposition 
by neutrinos in baryon inhomogeneities after the quark hadron transition.
If the total energy deposited by the neutrinos 
is $\Delta E$ and the 
volume increase  due to the energy deposition is $\Delta V$  then, 
\begin{equation}
\Delta V = {\Delta E \over \rho_{rad}}
\end{equation}
where $\rho_{rad}$ is the energy density of the plasma in the radiation epoch. 
If $n_b$ be the baryon number density in the inhomogeneity and $n_{b0}$ be the
baryon number density of the background plasma then the overdensity can be defined 
as $\delta_n = {n_b \over n_{b0}}$. 
As baryon number is conserved, one can then obtain \cite{heckler},
\begin{equation}
{d \delta_n \over dt} = {-\delta_n \over \rho_{rad} V(t)} {dE \over dt}
\end{equation}
or, 
 \begin{equation}
{d \delta_n \over dt} = {-\delta_n \over \rho_{rad}} {d\epsilon \over dt}
\label{master}
\end{equation}
where ${d\epsilon \over dt}$ is the rate of energy density deposited in the 
inhomogeneity. Eq. \ref{master} gives the evolution of the inhomogeneity with 
time.  
In the absence of a magnetic field this has been calculated both in ref.
\cite{jedamzik} and \cite{heckler}. It depends on the neutrino flux
(${g_\nu \over 10} T^3$), the weak cross-section ($G_F^2 T^2$), the total 
number of targets in the fluctuation and the average energy transfer during 
the collisions. Combining all these we get,  
\begin{equation}
{d\epsilon \over dt} =  0.2 G_F^2 ({\delta T \over T}) T^9
\label{edwm}
\end{equation}
The factor of $0.2$ changes to $1.8$ when electron-positron annihilations
are taken into account \cite{jedamzik}. Hence we will take the factor to be
$1.8$. $G_F$ is the Fermi constant and 
${\delta T \over T}$ is the fractional temperature difference between the 
inhomogeneity and the background plasma. For inhomogeneities present after the  
quark hadron phase transition, it can be obtained 
by using the pressure equilibrium condition 
between the inhomogeneity and the plasma as, \cite{heckler}
\begin{equation}
{\delta T \over T} = {\delta_{n0} \eta_0 \over  g_{eff}}
\label{deltatbyt}
\end{equation}
where $\delta_{n0}$ is the initial overdensity, $\eta_0$ is the baryon to photon
ratio of the plasma at that temperature and $g_{eff}$ is the effective degrees
of freedom  taken to be 10.75 here. 
The energy density deposited in the baryon inhomogeneitites due to the neutrinos            
depends on $\chi {\delta T \over \delta r}$ where $\chi$ is the heat
conductivity in an imperfect fluid.
$\chi$ depends upon the energy density carried by the neutrinos, the neutrino
mean free path and the temperature of the plasma. $ \chi = {4 \over 3}
{\rho_{\nu} \over T} \lambda_{\nu}$.
$\delta r$ is the size of the inhomogeneity. In our calculations we consider it
to be the same as the neutrino mean free path. Therefore,
\begin{equation}
\epsilon =  {4 \over 3}{\rho_{\nu} \over T} \lambda_{\nu}{\delta T \over \delta r}
\end{equation}
If $Q$ (the emmissivity) gives the rate of
energy density carried by the neutrinos ($ d\rho_{\nu} \over dt $ )from the
plasma, we can obtain the rate of heat deposition in the inhomogeneity as,
\begin{equation}
{d\epsilon \over dt} = {4 \over 3} Q {\delta T \over T}
\label{ede+e-}
\end{equation}
One thing to keep in mind over here is that the above expression is only for
$\delta r \sim \lambda_{\nu}$; for $\delta r \neq \lambda_{\nu}$  
there will be a factor multiplying
it which depends on the ratio of the neutrino mean free path to the size of the
baryon inhomogeneity. Once the value of $Q$ is obtained, we can obtain 
the energy deposited.

\section{Neutrinos in magnetic fields.}

Magnetic fields of different magneitudes have been predicted in the early 
universe \cite{kronberg}. Though neutrinos are uncharged, 
a magneitzed plasma nevertheless 
affects their interactions. Since we are interested in the heat lost by the
background plasma we concentrate on the $\nu-\bar{\nu}$ production from e+e- 
annihilation and the neutrino synchrotron radiation in the presence of a 
magnetic field. Though there are many calculations of these interactions
\cite{others}, we follow the analysis done by Kaminker et al. 
\cite{kaminker1,kaminker2}. This is because they have done the analysis 
for the largest range of temperatures. Since the temperatures we are dealing
with are higher than the temperatures commonly encountered in the neutron 
stars we find that only the analysis of ref. \cite{kaminker1,kaminker2} 
are applicable in our case. For $T > 10^{11} K$ and for high magnetic fields the 
rate of neutrino energy density carried by the 
$\nu-\bar{\nu}$ pairs per unit volume, has been obtained in ref. \cite{kaminker1}
and the energy density carried by the neutrinos in synchrotron radiation
has been obtained in ref.\cite{kaminker2}. The plasma in our case is 
non-degenerate and relativistic. We see that for such a plasma, the analysis of 
Kaminker et al show that the dominant heat loss mechanism from the plasma 
depends upon whether the magneitc field chosen is quantizing or not. 
The neutrino emmissivity calculation depends on the number density of plasma ions
occupying the Landau levels at a particular temperature. For a non-quantizing 
magnetic field many Landau levels are occupied, while for a quantizing magnetic
field the particles mostly occupy the ground 
level. Thus the majority of plasma particles are incapable of emitting 
synchrotron 
$\nu-\bar{\nu}$ pairs for quantizing magnetic fields. This is reflected
 in the presence of an exponentially small factor in the 
final expression of the synchrotron loss rate which
reduces the emmissivity considerably. 
Hence for a quantizing magnetic field, neutrino emmissivity 
from e+e- annihilations dominates over the emmisivity from synchroton 
radiation. 
The value of $Q$ for different temperatures and magnetic field have  
been calculated by Kaminker et al. for both $\nu-\bar{\nu}$ pair production and 
neutrino synchrotron radiation. We substitute these values in the equation
for energy deposition obtained in the previous section and study 
the evolution of baryon inhomogeneities due to neutrino inflation in 
the presence of magnetic fields. 

\section{Evolution of baryon inhomogeneities in the presence of a magnetic 
field}

The inhomogeneities are mostly formed just after the quark-hadron transition which
takes place around $170$ MeV. Hence we will consider the evolution of the 
inhomogeneities around this time.  Depending
on the magnitude of the temperature the magnetic field is either 
non-quantizing or quantizing.  The two cases have to be treated in different ways.
We take $ T \sim  100 $MeV.
So, the non-quantizing magnetic field will mean 
$B < 4.414 \times 10^{19} $Gauss, while the quantizing fields will be,
$B > 4.414 \times 10^{19} $Gauss. Since such high magnetic fields are difficult
to come by in the early universe, we can safely assume that for temperatures 
around the quark-hadron transition,
the background magnetic fields are non-quantizing. The emmissivity $Q$ is then
obtained for the relevent parameters from refs.\cite{kaminker1,kaminker2}.

Once $Q$ is obtained we substitute it in eqn.\ref{ede+e-}. Using eqn 
\ref{deltatbyt}, and the time-temperature relation,
\begin{equation}
t = (0.3 g_{eff}^{1/2}){m_{pl} \over T^2}
\label{tT}
\end{equation}
 in eqn.\ref{master}, we get, 
\begin{equation}
{d\delta_n \over \delta_n } = K {dT \over T^7}\left[{d\epsilon \over dt}\right]_{total}
\label{total}
\end{equation}
where $K = {1.8 m_{pl} \over (\pi^2 g_{eff}^{1/2})} \sim 10^{21} $ MeV and 
$[{d\epsilon \over dt}]_{total}$ is given by, 
\begin{equation}
\left[{d\epsilon \over dt}\right]_{total}=\left[{d\epsilon \over dt}\right]_{thermal}+\left[{d\epsilon \over dt}\right]_{e+e-}+\left[{d\epsilon \over dt}\right]_{syn}
\end{equation}
The first term on the R.H.S gives the energy deposited in the absence of a 
magnetic field. 
The second term on the R.H.S gives the energy deposited due to neutrino pair 
production from 
e+e- annihilation and the third term on the R.H.S gives the energy deposited due 
to neutrino synchrotron radiation. 
Integrating both sides of eqn.\ref{total}, we  find out how the inhomogeneity
evolves between a certain temperature range. We consider the initial
overdensity in the inhomogeneity to be ${\delta_{n0}} \sim 10^4$. 
We also make the approximation that ${\delta T \over T} $ is constant. 
For a more through analysis one must 
do a proper simulation where ${\delta T \over T} $  changes continuously. 

We integrate the R.H.S. of eqn \ref{total} separately for the three different 
mechanisms and join them at the end to get the final expression. Let, 
\begin{equation}
I_1 = \int[{dT \over T^7} \times C_1 T^9]
\end{equation}
where  $C_1 = 1.8 G_F^2 \delta_{n0}{\eta_{0}\over g_{eff}} $. Hence, 
\begin{equation}
I_1 = {C_1 \over 3}(T_2^3 - T_1^3)
\label{firterm}
\end{equation}
where we have considered $T_1$ as the initial and $T_2$ as the final 
temperature.  

For the second term, we see from ref.\cite{kaminker1}, for high temperatures
and non-quantizing magnetic fields, the emissivity of neutrino pairs due to 
e+e- annihilation is independent of the 
magnitude of the magnetic field and is given by, 
\begin{equation} 
Q = {7 Q_c \over 12 \pi} \zeta(5) \left({T \over 5.93 \times 10^9 K}\right)^9  (C_v^2 + C_A^2)
\label{e+e-lo}
\end{equation}
where $Q_c = 1.015 \times 10^{23}  erg cm^{-3} sec^{-1}$  and $(C_v^2 + C_A^2) = 1.675$.
Thus the second term will be similar to the first one after integration and will only have a different constant $C_2$. Therefore, 
\begin{equation}
I_2= {C_2\over 3}(T_2^3 - T_1^3)
\label{secterm}
\end{equation}
where $C_2 ={7 Q_c \over 12 \pi} \zeta(5) ({1 \over 5.93 \times 10^9 K})^9 (C_v^2 + C_A^2)\delta_{n0}{\eta_{0}\over g_{eff}} $. 
For the synchrotron radiation, the emmissivity in the high temperature,
non-quantizing magnetic field is given by \cite{kaminker2}, 
\begin{equation} 
Q = {20 Q_c \over 9 (2\pi)^5} \zeta(5) \left({T \over 5.93 \times 10^9 K}\right)^5 b^2 C_+^2
\left[ln\left({T \over \sqrt{b}\times 5.93 \times 10^9 K}\right)+ 2.33\right]
\label{synlo}  
\end{equation}
where $ b = {\mbox{B (Magnetic field in Gauss)} \over  4.414 \times 10^{13} \mbox {Gauss}}$ 
and $C_+^2 = 1.68$.
Therefore, 
\begin{equation}
I_3 = \int\left[{dT \over T^7} \times C_3 T^5\left[ln\left({T \over C_4}\right)+2.33\right]\right]
\end{equation}
where $C_3 ={20 Q_c \over 9 (2\pi)^5} \zeta(5) ({1 \over 5.93 \times 10^9 K})^5 b^2 C_+^2 \delta_{n0}{\eta_{0}\over g_{eff}}$ 
and $ C_4 ={1 \over \sqrt{b}\times 5.93 \times 10^9 K}$ are constants depending on 
the magnetic field. Integrating, we get, 
\begin{equation}
I_3 = C_3 [{ln({T_1 \over C_4}){1 \over T_1} - ln({T_2 \over C_4}){1 \over T_2}}+(C_4 + 2.33)({1\over  T_1} - {1 \over T_2})]
\end{equation}
So finally we have, 
\begin{equation}
\delta_n = \delta_{n0} Exp[{K \over 3}(C_1 + C_2)(T_2^3 - T_1^3) + C_3[{ln({T_1 \over C_4}){1 \over T_1} - ln({T_2 \over C_4}){1 \over T_2}}+(C_4 + 2.33)({1\over  T_1} - {1 \over T_2})]]
\label{lowmag}
\end{equation} 
We evaluate the constants and plot the change in $\delta_n$ 
as a function of 
temperature. Constants $C_1 $ and $C_2$ are independent of the magnetic field. 
We evaluate $C_3$ and $C_4$ for different values of magnetic fields. $T_1$ is 
taken to be  $150$ MeV, since we assume that the 
baryon inhomogeneities have been formed during the quark hadron transition which
takes place around $170$ MeV. The final temperature is taken to be about $100$ 
MeV. We have taken this because below $100$ MeV, our definition of quantizing
and non-quantizing magnetic fields may change. This would make our analysis
inconsistent.  

We find that for low values of magnetic fields there is not much difference. But
as we increase the value of magnetic field the difference starts increasing. 
For very high magnetic fields, $ B = 10^{18}G $ we see a very significant 
difference in the evolution of the inhomogeneity. This is shown in Fig.1. The
solid line denotes the evolution of the inhomogenity in the absence of a magnetic
field while the dashed line denotes the evolution in the presence of the magnetic
field. Clearly the inhomogenities get wiped out faster in the presence 
of a magnetic field. Of course, the magnetic field considered is very high, but
we will discuss later on the possibility of such high magnetic fields being
present in the early universe. Even for $ B = 10^{17}$ G, 
the evolution is not exactly the same.  For lower values of the magnetic field, 
the difference in the
evolution is still smaller, but there is always a finite difference with the
field free case. 
   
\begin{figure}
\begin{center}
\includegraphics{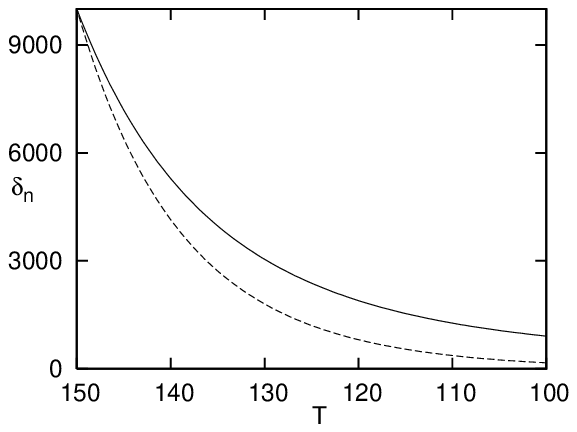}
\end{center}
\vskip -0.5cm
\caption{}{This figure shows the decrease in the overdensity of the inhomogeneity
over the temperature range 150 MeV - 100 MeV. The solid line denotes the 
decrease in the ovedensity in the absence of the magnetic field while the dashed 
line shows it for a magnetic field of magnitude $ B = 10^{18}G $.}
\label{Fig.1}
\end{figure}


All this is when the temperature is above $100$ MeV. However, neutrinos decouple
only after $1$ MeV. So these effects will also be there below $100$ MeV. Only
now the non-quantizing fields will have to be below
$ 4.414 \times 10^{17}$Gauss. The effect here is similar as before, as the field
is increased the inhomogenity decreases faster. 

We now do a similar analysis for quantizing fields.  
Here the emissivity from the e+e- annihilation is given by \cite{kaminker1},
 
\begin{equation} 
Q = {Q_c \over 48 \pi^3} b \zeta(3) \left({T \over 5.93 \times 10^9 K}\right)^5  (C_v^2 + C_A^2).
\label{hmee}  
\end{equation}
So that we get,
\begin{equation}
I_2 = \int[{dT \over T^7} \times C_2 T^5]
\end{equation}
where $C_2 ={Q_c \over 48 \pi^3} b \zeta(3) ({1\over 5.93 \times 10^9 K})^5 (C_v^2 + C_A^2) \delta_{n0}{\eta_{0}\over g_{eff}}$
 and is therefore dependent on the magnetic field.  Hence, 
\begin{equation}
I_2 = C_2( {1\over T_1} - {1\over T_2})
\end{equation} 
The emissivity from the synchrotron radiation in this case is given by, 
\cite{kaminker2}
\begin{equation} 
Q = {16 Q_c \over 9 (2\pi)^5}\left({T \over 5.93 \times 10^9 K}\right)^{1/2} b^4 C_+^2 (1 - {9 \over 4 e})(2 \pi \sqrt{2b})^{1/2}Exp[-{ \sqrt{2 b}\times 5.93 \times 10^9 K \over T}]
\label{synhi}  
\end{equation}
As pointed out in ref. \cite{kaminker2}, this is very small due to the presence 
of the exponential term. When we carry out the integration we get, 
\begin{eqnarray}
\nonumber I_3 = C_3[{Exp(-{C_4\over T}) [{1 \over (C_4 T^{9/2})}+{9 \over (2 C_4^2 T^{7/2})}+{63 \over (C_4^3 T^{5/2})}+{315 \over (8 C_4^4 T^{3/2})}+ {945 \over (C_4^5 T^{1/2})}]}] \\
\noindent { - C_3[945 \sqrt{\pi} Erf [{C_4 \over T}]^{1/2} ({1 \over (32 C_4^{11/2})}) ]}
\label{eqnerf}
\end{eqnarray}
where $C_3 = {16 Q_c \over 9 (2\pi)^5}({1 \over 5.93 \times 10^9 K})^{1/2} b^4 C_+^2 (1 - {9 \over 4 e})(2 \pi \sqrt{2b})^{1/2} \delta_{n0}{\eta_{0}\over g_{eff}}$ and
 $C_4 = \sqrt{2 b}\times 5.93 \times 10^9 K $.
Since as pointed out earlier, the synchrotron radiation does not
contribute much to the overall emmissivity in this region, hence the total 
decrease 
in the inhomogenity is less in the case of the quantizing magnetic 
field compared to the decrease in the non-quantizing case.  
Here only $C_1$ is independent of the magnetic field. So we have to obtain 
$C_2,C_3$ and $C_4$ for different values of magnetic fields. 
The results are given in fig 2. We see that (as observed in the previous 
case also ) as the magnetic field is 
enhanced, the inhomogeniety decreases faster.The solid line denotes the 
evolution in the absence of magnetic fields. The dashed line indicates the 
evolution of the inhomogeneity at a magnetic field of $ B = 10^{18}G $, while
the dotted line shows the evolution at $ B = 10^{19}G $. 
This decrease is less than in the previous case (fig 1), which seems to show that
it is the synchrotron radiation which plays a greater role in the heat loss 
mechanism than the neutrino production from e+e- annihilation.

\begin{figure}
\begin{center}
\includegraphics{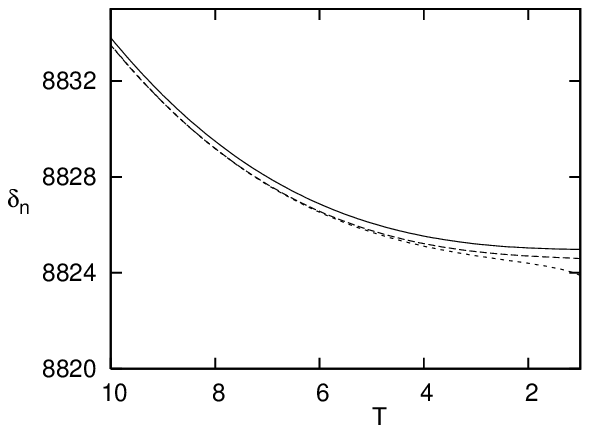}
\end{center}
\vskip -0.5cm
\caption{}{This figure shows the decrease in the overdensity of the inhomogeneity
in  the temperature range 10 MeV - 1 MeV. The solid line denotes the 
decrease in the ovedensity in the absence of the magnetic field, the dashed 
line shows it for a magnetic field of magnitude $ B = 10^{18}G $ and the 
dotted line shows the evolution at $ B = 10^{19}G $. The initial 
value of the overdensity at T = 100 MeV is taken to be $10^{4}$ }
\label{Fig.2}
\end{figure}

\section{High magnetic fields in the early universe.}
A detailed review of magnetic fields in the early universe is given in
ref. \cite{kronberg}. Here we mention only some special cases where very 
high field values have been postulated. Superconducting strings in the early 
universe may generate very high fields. In ref.\cite{berezensky}, the authors
have discussed the generation of magnetic fields with values as high as 
$10^{22}$
Gauss. Superconducting strings passing through the hadronic plasma after
the quark-hadron transition, thus 
may generate very high magnetic fields over large lengthscales. 
Large magnetic fields 
have also been postulated to explain extragalactic 
gamma ray bursts. 
 
Apart from this, since after the quark-hadron transition the magnetic 
field evolves according to the frozen-in law \cite{kronberg}, 
 calculation shows that if the magnetic field is produced at the 
quark hadron transition with maximum helicity and there was equipartition of
thermal and magnetic energy, then   
a magnetic field of magnitude $10^{17}$ Gauss on the scale of $30$ kms 
immediately after the phase transition is not implausable. Such large magnetic 
fields have been predicted in various models where the magnetic fields 
are generated by shock waves \cite{sigl}. Hence it is possible to have large 
magnetic fields after the quark hadron transition 
which will affect the neutrino inflation of baryon inhomogeneities.

\section{Conclusions and Discussions}
In conclusion we have established that strong magnetic fields do affect the 
inflation of baryon inhomogeneities by neutrino heating. We have seen that this 
effect is greater at higher temperatures and for non-quantizing fields. Our 
results seem to indicate that for the same value of the field, a non-quantizing
field has more effect than a quantizing field.  
This is expected, since as mentioned before, for the quantizing field, most 
of the plasma particles are in the ground Landau level and therefore are 
incapable of emitting synchrotron radiation. This reduces the number of neutrinos
depositing heat in the inhomogenities. 
Now,  whether the field would 
be quantizing or not depends on the temperature. Since at higher temperatures, 
even strong magnetic fields are non-quantizing, hence the effect described here
is greater at higher temperatures and stronger magnetic fields. 

We have simplified our 
calculations using certain approximations. The temperature difference
between the inhomogeneity is considered to be constant. However the temperature 
difference is actually related to the amplitude of the inhomogeneity and changes 
accordingly with it. But this change is very small. A proper investigation would 
involve a detailed simulation where the small temperature change should be 
taken into account.  The result of keeping this constant is that 
we get the largest possible effect. As the amplitude of the inhomogeneity decreases, 
the temperature difference also decreases but as our results show the 
inhomogeneity does not decrease very rapidly so we feel that our results will not 
change very much even if we take the small temperature changes into account. 
We have also taken the size of the inhomogeneities equal to the mean free path of neutrinos 
at that temperature. If the inhomogeneities are larger then the neutrinos cannot penetrate the 
entire inhomogeneity and will affect only the edges of the inhomogeneities. However since neutrino
mean free path increases as temperature decreases, most of the inhomogeneities within the horizon 
will have their sizes either equal to or smaller than the neutrino mean free path. For smaller 
size inhomogeneities, the heat deposited will be less by a factor given by the ratio of the 
size of the inhomogeneity to the mean free path of the neutrinos at that temperature.

We have considered baryon 
inhomogenities present after the quark-hadron transition but there is the
possibility of inhomogenities being present much before that. Since these 
inhomogenities may be created anytime after the electroweak phase transition,
a much through investigation of their evolution should include the high 
temperature zone ($100$ GeV -$200$ MeV) also. For this one has to study 
neutrino interaction properties at much  higher temperatures, in the presence
of a magnetic field. To our knowledge, such a study has not been carried out. 
We hope that this work will provide a motivation to study neutrino emission
in a magnetic field at GeV temperatures and its subsequent effect on the 
early universe.

\vskip .2in
\centerline {\bf ACKNOWLEDGMENTS}
\vskip .1in
I would like to thank Ajit Srivastava, Biswanath Layek and Kaushik Bhattacharya
for their suggestions and comments. I would also like to thank Charles Horowitz, 
Adam Szczepaniak and the Nuclear Theory Center, Indiana University, Bloomington
for their support and hospitality.

\end{document}